\documentclass[prd,nofootinbib,twocolumn]{revtex4}
\usepackage[usenames,dvipsnames]{color}
\usepackage{ifpdf}
\ifpdf
  \usepackage{feynmf} 
\else
   \usepackage{feynmp} 
\fi
\usepackage{graphicx}
\usepackage{amsmath}
\usepackage{amsfonts}
\usepackage{amssymb}
\usepackage{dsfont}
\usepackage{dcolumn}
\usepackage{bm}
\usepackage{slashed}
\usepackage{pstricks}
\usepackage{slashed}
\usepackage{multirow}
\usepackage{array}
\usepackage{rotating}
\usepackage{xcolor}
\usepackage{epstopdf}
\usepackage{fancybox}
\newcommand{\Usq}{ {\bar m}^2_{H_{u}}  }

\newcommand{\Tsq}{ m^2_{T}   }
\newcommand{\Qsq}{ m^2_{Q_3}  }

\newcolumntype{x}[1]{
>{\centering}p{#1}}%

\newcommand{\GeV}      {~\mathrm{GeV}}

\newcommand{\beqn}{\begin{eqnarray}}
\newcommand{\eeqn}{\end{eqnarray}}
\newcommand{\ba}{\begin{eqnarray}}
\newcommand{\ea}{\end{eqnarray}}
\newcommand{\be}{\begin{equation}}
\newcommand{\ee}{\end{equation}}

\newcommand{\mathsym}[1]{{}}

\def \n34{\tilde{\chi}^{0}_{3,4}}

\def \ta{\tilde{t}_1}
\def \tb{\tilde{t}_2}
\def \ba{\tilde{b}_1}

\def\met100{\slashed{E}_T\geq 100 \GeV}

\usepackage{ulem,fancyvrb}

\def\met{\slashed{E}_{T}}

\begin{document}

\title{A NEW (STRING MOTIVATED) APPROACH  \\ TO THE LITTLE HIERARCHY PROBLEM }
\author{Daniel Feldman,  Gordon Kane, Eric Kuflik, and Ran Lu}
\affiliation{Michigan Center for Theoretical Physics, University of Michigan, Ann Arbor,
MI 48109, USA. }



\begin{abstract}
{
We point out that in theories  where the gravitino mass,  $M_{3/2}$, is in the range (10-50)~TeV, with soft-breaking scalar masses and trilinear couplings of the same order, there exists a  robust region of  parameter space where the conditions for electroweak symmetry breaking (EWSB) are satisfied without large imposed cancellations. Compactified string/M-theory with stabilized moduli that satisfy cosmological constraints generically require a gravitino mass greater than about 30~TeV and provide the natural explanation for this phenomenon. We find that  even though scalar masses and trilinear couplings (and the soft-breaking $B$ parameter) are of order (10-50)~TeV, the  Higgs vev takes its expected value and   the $\mu $ parameter is naturally of order a TeV. The mechanism provides a natural solution to the cosmological moduli and gravitino problems with EWSB.
}

\end{abstract}

\maketitle

\section{Introduction and Motivation}

There is a serious problem in particle physics, called the `little hierarchy' or `fine-tuning' problem. The issue is that if we can explain the value of the $Z$ or $W$ boson masses, or equivalently the Higgs boson vacuum expectation value (vev), then these quantities have to be calculated in terms of new physics scales. The Higgs vev cannot be derived in the Standard Model itself. But direct and indirect lower limits on masses of new particles are large enough that any
explanation must involve large cancellations or suppressions that seem fine-tuned, typically by one to two orders of magnitude.  The problem is often stated in terms of supersymmetric models, but it is equally serious for all approaches to breaking the electroweak symmetry, because all approaches require heavy new particles.  Sometimes the fine-tuning issues are described in the MSSM in terms of both the small $M_Z$ and the Higgs mass. The stop masses and mixing that determine the one loop correction to the Higgs mass tend to be over constrained. But this is a special problem in the MSSM only. If the gauge group is extended the tree level Higgs mass can get a significant contribution from D-terms, and other new physics can increase the Higgs mass. But the $Z$ mass, or the Higgs vev, are always small, so we focus here on the $Z$ mass and Higgs vev. Here we address this issue  in a supersymmetric framework, motivated by progress  which has been made in models based on compactified string theory, where the supersymmetry is softly broken and the soft-breaking Lagrangian is derived at a high scale, and then the electroweak symmetry is broken by the usual radiative mechanism with the minimal supersymmetric  field content~\cite{Everett}.

We will describe here how compactified string/M-theory suggests a new approach to the little hierarchy problem, but the approach is valid in any theory with heavy scalars and trilinear couplings of similar magnitude.  Compactified string theories have moduli that parameterize the shapes and sizes ot the curled up small dimensions.  One can show that generically the lightest eigenvalue of the full moduli mass matrix, is less than or of order the gravitino mass.  For the complete derivation and the numerical factors see \cite{Kuflik}; see also  \cite{Quevedo,Denef:2004cf,Scrucca} for earlier work. The implications of tying together the light moduli masses and the gravitino mass are very important -- they cannot be chosen independently, and the moduli masses must obey cosmological constraints.

Moduli decay by universal gravitational operators, so their lifetimes can be calculated~\cite{MoroiRandall,Watson,KaneFeldman} and their decays grow as their mass cubed. To avoid interfering with nucleosynthesis the moduli must then be heavier than about 30 TeV, so in  realistic string compactifications  the gravitino mass must also be heavier than about 30 TeV. From supergravity calculations, that in turn implies that \textit{scalar}  superpartners (squarks, sleptons) must be heavier than about 30 TeV (and will not be produced at LHC). However, in string models the gauginos normally can be much lighter and their signatures at the LHC should be present.

Having scalar superpartners so heavy seems to imply that the explanation of electroweak symmetry breaking and the small Higgs vev leads to a severe little hierarchy problem.  For some early speculations on very heavy scalars, but otherwise different from our approach, see \cite{JW}.  On the other hand, if the results come from an underlying 10 or 11 dimensional string/M-theory, and if EWSB occurs robustly in the theory, then there is essentially by definition no hierarchy problem. The results would follow not by fine-tuning but inevitably from the underlying theory. 

Motivated by such thinking, we have found the mechanism that allows an apparent cancellation to occur and to explain a   small higgs vev and $\mu$,
the effective higgsino mass parameter. It is different from previous approaches to fine-tuning and arises from a different region of parameter space than previous approaches. The solution is general and gives a robust region where EWSB occurs in generic string motivated theories that satisfy the above requirements. In particular, the supergravity formalism implies that the trilinear couplings also are given by the gravitino mass with a coefficient of order unity, and maintaining this is a key aspect of obtaining EWSB robustly without introducing what would naively be expected to be an enormous tuning.

In what follows we write the EWSB conditions and show how they can be satisfied with greatly reduced fine-tuning.

\section{General Mechanism}

Now we describe in some detail a generic solution to electroweak symmetry breaking  in supergravity and string motivated models that  gives rise to a new approach to the little hierarchy  problem. In the analysis, we will use a common scalar mass ${M}_{0}$ and a common trilinear ${A}_{0}$, with ${M}_{0}\simeq {A}_{0} \simeq M_{3/2} \simeq 30 \rm  ~\rm TeV$, which naturally arises in string compactifications. As mentioned in the introduction we expect the following mass hierarchy 
\be M^2_0 ,  A^2_0  , {B}^2_0    \gg  \mu^2, M^2_{a} ~, \ee  
where $a$ indexes the gauginos of $SU(3), SU(2),U(1)$, (i.e. gluino, wino, bino soft masses) which at the unification scale will generally be split,  and the index 0 indicates unification scale values, and $B = B_{\mu}/\mu$. By ${A}_{0}^{2}$ we will always mean magnitude $|{A}_{0}|^{2}$ throughout.

\begin{figure}[t!]
\includegraphics[scale=0.4]{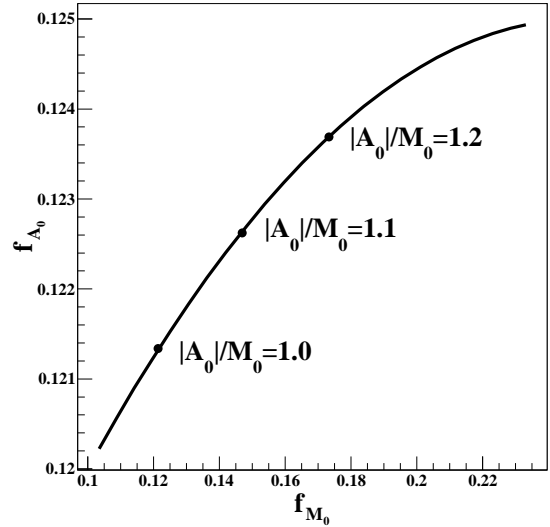}
\caption{\label{f1f2} 
The 1 loop RGE coefficients ${ f}_{M_0}$ and ${f}_{A_0}$ at $Q_{\rm EWSB}$ as given in Eq.(\ref{ipeq}). 
The amount of cancellation in the  Eq.(\ref{ipeq}) for  $m_{H_u}^2(Q_{\rm EWSB})$ depends on $|A_0|/M_0$, 
and we show the values that minimize  $m_{H_u}^2$ at one-loop.  In this
figure, $M_0$ runs from 10 TeV at the lower end of the curve to 50 TeV at the top of the curve.
  (See Fig.~(\ref{sweep}) for the full  analysis with with 2 loop running and the threshold/radiative corrections.) }
\end{figure}

The RG equations for the Higgs mass-squared parameters $m_{H_{u}}^{2}(t)$ and $m_{H_{d}}^{2}(t)$--which will feed directly into the calculation of the Higgs vev (see Eq.(\ref{EWSBeqn1}))--shows that $m_{H_{d}}^{2}$ essentially does not run, while $m_{H_{u}}^{2}$ does, so that one has that $m_{H_{d}}^{2}(t) \simeq M^2_{0}$ and
\begin{equation}
m_{H_{u}}^{2}(t) \simeq f_{M_0}(t){M}_{0}^{2}-f_{A_0}(t){A}_{0}^{2} + R(t)~, \label{ipeq}
\end{equation}%
where $t=ln(Q/Q_{0})$, $Q_{0}$ being the unification scale.   The functions ${ f}_{M_0}, { f}_{A_0}$ are, to leading order, determined by Standard Model quantities (gauge couplings and Yukawa couplings)  and the unification scale.   Analytical formulas for ${ f}_{M_0}, { f}_{A_0}$ are given in the Appendices   for one-loop running, while the numerical calculations are performed using the full two-loop RG equations.   $R(t) = f_{3}(t) A_0 M_3(0)+ f_{4}(t) M^2_3(0)+ \ldots$ are corrections that are  smaller or the same size as the sum  of the first two terms in Eq.~(\ref{ipeq}). One finds that  ${ f}_{M_0} $ and  ${ f}_{A_0}$ at the EWSB scale  have a value of
\be
f_{M_0} \simeq f_{A_0} \simeq 0.1~~. \label{supr1}
\ee  Thus $m_{H_{u}}^{2}$
 is suppressed by the values of ${ f}_{M_0}$, ${ f}_{A_0}$.  To illustrate this effect, we plot $f_{M_0}$ vs. $f_{A_0}$ in Fig.~(\ref{f1f2})  which shows the dependence of $m_{H_u}^2(Q_{\rm EWSB})$ on the soft masses is reduced, as is the size of $m_{H_u}^2(Q_{\rm EWSB})$ relative to $M_{3/2}$.

At first it may seem that results in Eq.(\ref{supr1}) are independent of the scale of the soft masses, but in-fact large scalar masses, of order 10 TeV and larger, are necessary for this effect. This scale already arises as a bound from BBN on moduli-masses, which in turn gives a similar bound on the soft-breaking scalar superpartner masses~\cite{Kuflik}. As discussed in the Appendix~A, the values of ${ f}_{M_0} $ and ${ f}_{A_0}$ are mostly sensitive to the top Yukawa coupling. The top mass receives large $(10-15)\%$ corrections from squark/gluino loops~\cite{Pierce} in the models we discuss, resulting in a lower top Yukawa coupling required to produce the correct top-quark mass.  In other types of models which are not the type studied here, loop corrections from lighter squarks below  about $5 ~\rm TeV$ do not provide sufficient suppression, and the large Yukawa coupling would drive $m_{H_{u}}^{2}$ negative.

String motivated models predict an additional
 cancellation in $m_{H_{u}}^{2}(Q_{\mathrm{EWSB}})$, since the supergravity Lagrangian generically predicts that
\beqn
{M}_{0} \simeq  {A}_{0} \simeq M_{3/2}~.
\eeqn
Since the values of ${ f}_{M_0}$, ${ f}_{A_0}$ are naturally each  order $0.1$ and their difference leads to another suppression of order $0.1$,   the natural scale of $m^2_{H_u}(Q_{\rm EWSB})$ is 
\be
m_{H_u}^2 \sim 10^{-2} M_{3/2}^2 \sim \rm \mathcal{O}(TeV^2)~~. \label{double}
\ee

{\it Thus, the effects of the large $M^2_0$ and $A_0^2$ in the determination of  $m_{H_{u}}^{2}(Q_{\mathrm{EWSB}})$ are absent, and the naive fine-tuning is significantly reduced}. 

As a result of this cancellation the corrections of the size $R$  in  Eq.(\ref{ipeq}) are smaller or the same size  as
the term that cancels: $ f_{M_0}(t){M}_{0}^{2}-f_{A_0}(t){A}_{0}^{2}$.   This allows for  a value $\mu$ (and $M_3$) at the electroweak symmetry breaking scale that is of order a TeV or smaller when the soft scalars masses and trilinear couplings are large, in the range (10-30)~TeV or larger.  If we explicitly forbid a $\mu $ term in the superpotential (this can be done consistently~\cite{Ibanez,witten,akkl}), in which case supergravity implies that the soft breaking ${B_0}\approx 2 M_{3/2}$. Relaxing  this condition  a bit, we will find a reduced $\mu$ generally occurs close to  this prediction.

Now recall the two familiar EWSB conditions  
\beqn
\mu ^{2} &=&-{M}_{Z}^{2}/2+\frac{{\bar{m}}_{H_{d}}^{2}-{\bar{m}}_{H_{u}}^{2}\tan ^{2}\beta 
}{\tan ^{2}\beta -1 \label{EWSBeqn1}} \\
B\mu &=&\frac{1}{2}\sin 2\beta ({
\bar{m}}_{H_{u}}^{2}+{\bar{m}}_{H_{d}}^{2}+ 2 \mu ^{2})~,
\eeqn
where ${\bar m}_{H_{u}}^{2}$ includes the tadpole corrections to $m_{H_{u}}^{2}$. At the electroweak scale we can rewrite these with the approximations ${\bar m}_{H_{d}}^{2},B^{2}\gg {\bar m}_{H_{u}}^{2}$,  and not too small $\tan\beta$. Then $\sin
2\beta \approx 2/\tan \beta ,$ and $2B\mu \approx \sin 2\beta \bar{m}_{H_{d}}^{2}.$   In the above, ${\bar m}_{H_{d}}$ is essentially $M_{3/2}$ and  $B\approx 1.7M_{3/2}$ as summarized in Appendix~A. Combining these gives $\tan \beta \approx \bar{m}_{H_{d}}^{2}/B\mu$. Using this in the first EWSB condition gives a quadratic equation for $\mu ^{2}$, with an approximate solution (after some algebra),
\be
\mu^2 - \frac{M_{Z}^{2}}{2} \frac{{\bar m}_{H_{d}}^{2}}{B^{2}-{\bar m}_{H_{d}}^{2}}
\approx \frac{{\bar m}_{H_{d}}^{2}}{B^{2}-{\bar m}_{H_{d}}^{2}}{\bar m}_{H_{u}}^{2}~,
\label{mu}\ee
where the coefficient $\frac{{\bar m}_{H_{d}}^{2}}{B^{2}-{\bar m}_{H_{d}}^{2}} \simeq O(1/2)$.
\begin{figure}[t!]
\includegraphics[scale=0.42]{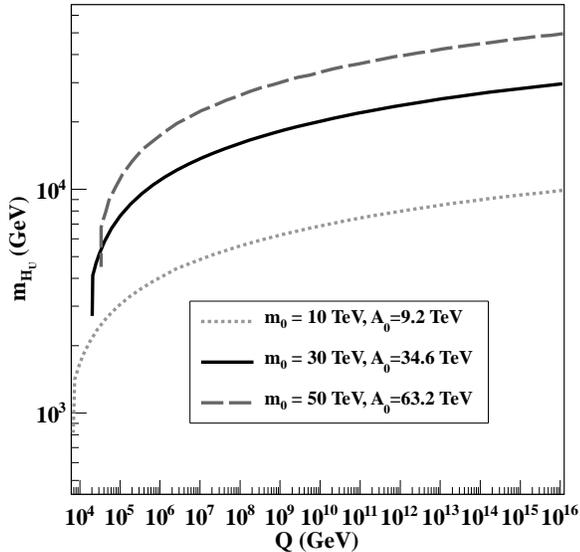}
\caption{\label{run} 
 Two-loop renormalization group running of  ${m}_{H_u}$ for 3 models
 for the cases  $M_0 = (10,30,50)~\rm TeV$.   
The tadpole corrections are shown, and appear as a vertical drop at  $Q_{\rm EWSB} = \sqrt{m_{\ta} m_{\tb}}$ as is appropriate.
The numerical value of ${\bar m}_{H_u}$, which is the tree + tadpole value,
continue  to take the same value at scales $Q$   below the point $Q_{\rm EWSB}$   
as is theoretically expected.
The values of $\mu$ are
$\mu =( \rm 500~GeV,
~1.0 ~TeV, ~1.8 ~TeV)$.  This can be seen for example for the $M_0 =30~ \rm TeV$  
 in figure \ref{sweep} using Eq.~(\ref{mu}). }
\end{figure}
The cancellation in Eqs.~(\ref{ipeq},\ref{supr1},\ref{double}) coupled 
with equation, Eq.~(\ref{mu}),
can be taken as our basic result.
Eqs.(\ref{ipeq},\ref{supr1},\ref{double},\ref{mu}) shows that with inputs
having all the soft-breaking parameters of order $30~\rm TeV$ one finds the
conditions for EWSB are always satisfied for reasonable ranges of the
parameters, and the values of $\mu$ can be at (or  below) the TeV scale
consistent with  EWSB and the measured value of $M_{Z}$.  

Equation~(\ref{mu}), with the important numerical values,
is realized naturally with heavy scalars  $M_0$ and large bilinear $B_0$ and trilinear $A_0$ couplings
of comparable size. 
We add that Eq.~(\ref{mu}) is a derived result and not assumed; we interpret
this as then a true prediction of string models with the breaking of superysmmetry
through gravitationally coupled moduli. 
We note that the result  $\mu^2 \sim {\bar m}_{H_{u}}^{2}$ can
be obtained in  gauge mediation (where $\mu^2 \ll B \mu$)  in models for which
$ B \mu \ll  {\bar m}_{H_{d}}^{2}$ is assumed~\cite{Csaki:2008sr} (see also \cite{Dine:1997qj}).
\begin{figure}[t!]
\includegraphics[scale=0.42]{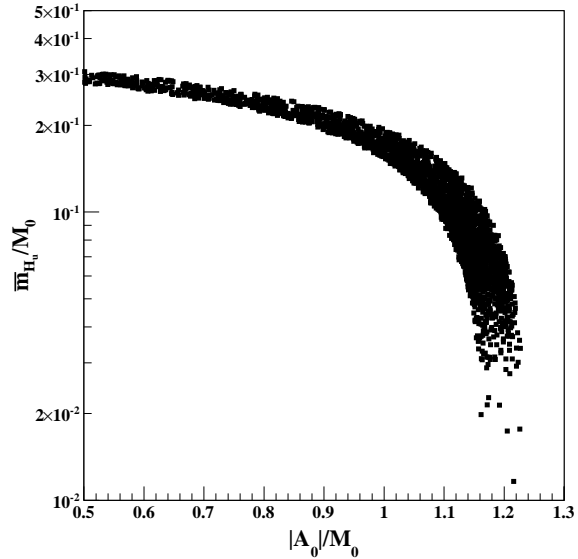}
\caption{\label{sweep} 
 A large parameter space sweep  using the full numerical analysis discussed in the text
  for $M_0 = 30 ~\rm TeV$, with $\mu \in [0.9,2]~\rm TeV$, with $\tan \beta \in [3,15]$
 showing a robust region where ${\bar m}_{H_u}$, the loop corrected
 value  at the EWSB scale,  is reduced significantly relative to $M_0 = 30 ~\rm TeV$,
with the greatest suppression occurring for trilinear of about the same magnitude. }
\end{figure}

In our numerical analysis we employ the 2-loop renormalization group equations (RGEs) for the soft supersymmetry breaking masses and couplings \cite{Martin:1993zk} with radiative corrections to the gauge and Yukawa couplings as computed in \cite{Pierce}. Radiative electroweak symmetry breaking is carried out with SOFTSUSY~\cite{Allanach}. In the Higgs sector, we include all the 2-loop corrections \cite{Zwirner,Slavich}. Explicitly we find Eq.~(\ref{mu}) is a consistent representation of $\mu$ for 
\be M_{3/2} = M_0 = 30~ \rm TeV~{\rm with}~ \mu \in [0.9, 2] ~\rm TeV ~.\ee
For $M_{3/2} = M_0 = 10~\rm TeV$ we find $\mu$ as low  300 GeV, though about (500-600)~GeV  appears more `natural' as a lower limit from our scan of the parameter space. 
In the numerical analysis we increase the maximum trials 
 in obtaining the solution of the RGEs relative to the default number in SOFTSUSY which serves, in part, to optimize our focussed scan.
This is described in more detail in Ref.~\cite{Allanach}. 
Our analysis is focussed on   $M_0  \in [10,30]~\rm TeV$, with $M_0 \simeq |A_0|$. We do not perform an extensive study of solutions with $M_{3/2} \gtrsim 50~\rm TeV$ because the programs may not be reliable there with the desired accuracy.

Figure~(\ref{run}) shows  $m_{H_{u}}$ for appropriate $A_{0},M_{0}$, 
 and how it runs down to values of order $M_{3/2}/10$ from the RGE effects alone, 
where in the last step the  Coleman-Weinberg corrections to the potential brings down the size of the up type Higgs mass$^2$ soft term   by additional factors. It is the very fact that string models tell
us ${B_0}\sim  M_{3/2},{A_0 }\simeq M_{3/2}, M_0 \simeq M_{3/2}$ that leads to this solution. If one put the trilinear coupling to zero, this solution to a very  large reduction of $\mu$  would be missed.  The result we propose here is very different from the focus point solution where $m_{H_{u}}^{2}$ runs to a common invariant value and turns tachyonic~\cite{Arnowitt:97,Nath:97,Feng:2000}, and for which  scalar masses and  trilinear couplings order 30 TeV were not discussed. We elaborate further on this in the Appendices.

 In contrast we discuss here a new phenomenon where there is a cancellation of the coefficients of $A_{0}^{2}$ and $M_{0}^{2}$ which are of comparable size but with opposite sign  and thus results  in a suppression of $m_{H_{u}}^{2}$ relative to the very heavy gravitino mass  of order $(10 - 50)~\rm TeV$. The solution for $\mu$ in the models we discuss is naturally in the range $\mu \lesssim (0.5-3)$ TeV  for $M_0 = (10 - 50) ~\rm TeV $ when $|A_0| \simeq M_0$. Differently,  in our case one can think in terms of a cancellation of the two contributions to Eq.(\ref{ipeq}), but it is natural  and not tuned.  The
top Yukawa runs significantly from the GUT scale and drives the soft up type scalar higgs mass parameter to be small relative to the gravitino mass and it is positive and not tachyonic, and in   addition, the large trilinear also leads to a faster running of $m^2_{H_{u}}$.

As mentioned above, Fig.~(\ref{run})  shows the the running of the up type Higgs soft mass for 3 sample  models with values of $\mu$ at EWSB scale,  $\mu =( \rm 500~GeV, ~1.0 ~TeV, ~1.8 ~TeV)$, for the three cases  $M_0 = (10~\rm TeV,30~\rm TeV,50~\rm TeV)$ and with the other soft breaking parameters of comparable size, and
with the $SU(2), U(1)$ gaugino masses well below 1 TeV; the  gluino masses are the heaviest and range from $400~\rm GeV$ to $1~\rm TeV$.  In Figure~(\ref{sweep}) we show a robust and  large parameter space, for the case $M_0 = 30~\rm TeV$ where the largest suppression of the loop corrected value,  ${\bar m}_{H_{u}}$,  occurs for a trilinear coupling of size $M_0$ which in turn suppresses $\mu$ at the EWSB scale.  For the case of $M_0 = 30~\rm TeV$, one sees the largest suppression of ${\bar m}_{H_{u}}$ occurs at $|A_0|/M_0 \simeq 1.2$.  For the case of $M_0 = 10~\rm TeV$ we find a similar analysis shows the point of maximal suppression occurs closer to $|A_0|/M_0 \simeq 0.9$.

\section{Conclusion and Discussion}

We have found a new approach to the little hierarchy problem, which may be interpreted as its inevitable solution,  and that occurs generically in string models whose field theory limit posses $N=1,D=4$ supergravity and whose solution is consistent with cosmological constraints on the presence of moduli fields that couple to massive visible superpartner states. With heavy scalars, squarks, sleptons, trilinears, bilinear $B$ term, and moduli, which are all of order $M_{3/2}\simeq 30~\rm TeV$, the $\mu $ parameter and $M_{Z}$ can be small, supressed by a factor of order 30 or more relative to the gravitino mass. Essential to this solution is that $m_{0},A_{0}$ are both large though their ratio is order unity as expected in string-based models of the soft-breaking Lagrangian. In addition because $B_{0}$ is also of order $M_{3/2}$, such a solution arises naturally. We argue that the natural scale of $\mu $ is about $(1-2)~\mathrm{TeV}$ though smaller values are uncovered.

If this proposed situation describes nature, it adds to the motivation for expecting to observe the phenomenological implications of a superpartner spectrum with very heavy scalars and  sub TeV  gauginos at the LHC \cite{AtlasSUSY,CMSSUSY}, \cite{FKRN,KKRW} and in dark matter experiments~\cite{XENON100, WMAP, PAMELA}  .

Interpreting the phrase `fine-tuning' requires thought. Some people want it to mean that all superpartners (or other new particles in a different theory) are individually of order $M_Z$, so the EWSB condition (e.g. Eq.~(\ref{EWSBeqn1})) is satisfied without cancellations at all. Since existing limits on superpartner masses have for some time been too large for that to occur, it is unclear what goal such arguments have. Fine-tuning is unexpected in physics, and unnatural, so if someone says something is fine-tuned they must mean that some solution exists that is not fine-tuned.

Our approach suggests to us that this is not the best way to think about it. We find that in an underlying theory where the TeV scale emerges (string theory or any other), it is natural to have heavy scalars (many, many TeV) and to have predicted Higgs vevs (and $\mu$ the higgsino mass mixing parameter)  of order a TeV or even less, but the mechanism that ensures this seems unlikely to give values for these quantities an order of magnitude smaller than a TeV, which anyone who calls having TeV values `fine-tuned' would have to hope for.

This discussion suggests an interesting interpretation. On the one hand the hierarchy between the order 30 TeV gravitino and scalar masses and the sub-TeV to TeV scale is natural.  On the other hand, the results are valid for a range of small higgs vev and $\mu$. One can imagine that a range of small values of the Higgs vev could arise in generic string compactifications, with the actual value or some nearby value being equally valid. From first principles we could calculate the Higgs vev approximately but not exactly. Indeed, a study of the range of  values of the Higgs vev  \cite{Donoghue}  that seem not to change the phenomenology of our world concluded that a range of Higgs vevs was consistent with our world.  It seems worthwhile to pursue this question in particular compactifications.

Concluding, we simply remark  that this  suppression of $\mu$,  putting it into the TeV region relative to $M_{3/2}$, which  is on the  order of $(10-50) ~\rm TeV$, is a remarkable and  non-trivial phenomenon, occurring over a large region of parameter space in well motivated models (See e.g. Fig.~(\ref{sweep}) and Eq.~(\ref{mu}) or Eq.~(\ref{musp})). The above constitutes what may be interpreted as a consistent solution to cosmic moduli  problem as it leads to EWSB that is robust, and rather natural.
 
Some readers may prefer to interpret our results in an effective theory sense -- theories with the scalars, trilinears, etc. of size $M_{3/2}$ which are order 30~TeV with suppressed gaugino masses will not need large cancellations to obtain a small Higgs vev.  But it also important to understand that this result is a surprising and correct prediction of compactified string/M-theory with moduli having masses that are in accord with cosmology.

\appendix
\section{Illustration of the  [Intersection Point] Solution  with one-loop RGE Analysis}
While the numerical analysis presented here includes all 2 loop effects as discussed in the text, we will now proceed to show semi-analytically at the 1 loop level
that the  effect of driving $\mu$ low relative to the scalars and the trilinear coupling, whose mass scales are  order the gravitino mass, $M_{3/2}$, is in the running of  ${m}^2_{H_{u}}$ plus an internal cancellation. One can see this by solving the RG equations under the approximations discussed in the text, which we will exhibit for the case of universal scalars masses and trilinear couplings.
Solving for the running of the square of the top  yukawa coupling,  $h_t$, for the lower end of the $\tan \beta$ range we consider,  one has 
\beqn h_{t}(t) &=&  y^2_{top}(t)/16  /\pi^2  =  h_{t}(0)  E(t) \delta(t)  \\
\delta(t)  &=& (1-12 h_t(0) F(t))^{-1}, ~~ F(t) = \int^t_0 E(t') dt' \ ,  \nonumber\eeqn
where $t=ln(Q/Q_0)$ and $E(t)$  depends on the gauge couplings and the unification scale
\beqn E(t) = (1+6\tilde \alpha\cdot  t )^{-16/9}(1-2\tilde \alpha \cdot t)^{3}(1-66 \tilde \alpha/5 \cdot t)^{13/99 } \nonumber\\ \eeqn 
and where $\tilde \alpha = \alpha_0 /( 4\pi) $ and $\alpha_0$ is the the square of the unified  gauge coupling in units of $4 \pi$.  The above is well known~\cite{IbanezLopez}. Meanwhile, at 1 loop $B = B_0 +\frac{1}{2}(\delta(t)-1)A_0$.

 Next we present the solutions for the scalar masses under the same approximations above, they are:  
\beqn
m^2_{S}(t)  = b_S (( C_S \cdot  M^2) + H(t,0))~,
\eeqn
where the dot product above is for  $M^2 = (m^2_{H_u}(0),\Tsq(0), \Qsq (0))$,   with   $C_{{H_{u}}} =(1,-1,-1),$ $C_{T} =(-1,2,-1),$  $C_{Q_3}= (-1,-1,5)$,  and $b_S = (1/2,1/3,1/6)$. Here $H(t,0)$ is given by
\be
H(t,0) = \frac{ \sum_S m^2_{S}(0)}{1-12 h_t(0) F(t)} + \frac{12 h_t(0) F(t) A^2_t(0)}{(1-12 h_t(0) F(t))^2}~.
\ee
Explicitly for the case of universal scalars $m^2_{S}(t)  = m^2_{S}(0) = M^2_0$ and trilinear, $A_t(t) =A_t(0)\delta(t)$, and  one has (at l loop)
\beqn
 \label{IPexample}
m^2_{H_{u}}(t) & = & M^2_0 \left[ \frac{1}{2} (3 \delta(t)-1)\right]-A^2_0 \left[ \frac{1}{2}  (\delta(t)-\delta^2(t) ) \right] \nonumber  \\
f_{M_0}(t) & = & \frac{1}{2} (3 \delta(t)-1), ~~~ f_{A_0}(t)  =  \frac{1}{2} (\delta(t)-\delta^2(t))~, \nonumber  \\
\eeqn
 which gives $f_{M_0}$ and $f_{A_0}$ at the leading 1-loop level. This approximation above describes well the full result after including the tadpole corrections at the EWSB scale and the  corrections in the text arising from the products $R  \propto  M_{3}(0) A_0$ and $R \propto M^2_{3}(0)$.   

The top mass also receives important corrections from  top squark/gluino loops (see \cite{Pierce}).  The models discussed here have $\Delta m_{top} /m_{top} \sim (10 -15)\% $ for soft breaking scalar masses of size $M_0 \in [10,30]~\rm TeV$. The shift in the top Yukawa relies on the  corrections  computed in \cite{Pierce,Allanach}. The main effect we wish to emphasize from the above  is the large cancellation from $A_0 \simeq M_0 \simeq M_{3/2} \simeq (10-50)~\rm TeV$ via Eq.~(\ref{IPexample}) which drives $m^2_{H_{u}}$ small (but positive) when the gauginos are much suppressed.

We refer to this approach to the hierarchy  as an \underline{\it Intersection Point}  (IP), for  it is this cancellation in Eq.(\ref{IPexample}) involving $\delta(t)$, or near intersection of the 2 terms in square brackets, each positive and each order $1/10$ that drives $\mu$ small. Residual corrections to the right hand side, from the product $R$    will shift the IP and the complete solution has these corrections, but they are small for the models we discuss. Putting $M^2_0=A^2_0 = M^2_{3/2}$  for the case when $R$ can can be completely ignored (for very light gluino mass with scalars very heavy), remarkably the minimum is 
$ \delta(t)_{min}=-1+\sqrt{2}$  - in general the corrections via $R$ are present. One finds actually from the running that large suppressions can occur with solutions $\delta_{true}$ very close to $\delta(t)_{min}$; analytically and numerically when $A_0 \simeq M_0 \simeq 30 ~\rm TeV$ one obtains
at the breaking scale
\be
\mu^2 \sim \frac{1}{2} \Usq = {\cal O}\left(\frac{1}{10^{2}}\right)M^2_{3/2}  \label{musp}
\ee
where lowers value are possible, but slightly less natural.
In the above,   $\bar m^2_{H_i} = m^2_{H_i} - T_i/v_i $ , where $\Sigma_i = -T_i/v_i$ (see e.g.  \cite{Slavich}) are the tadpole $i=(d,u)$ corrections. We note in passing that  one can check at each order in the loop corrections to EWSB~\cite{Slavich},  and at each loop order in the determination of the  mass of the light cp-even higgs~\cite{Zwirner,Martin:2007pg} (whose mass is in the range  $(114-135)~\rm GeV$  in the models we discuss), that the suppression of  
 $\sin 2 \theta_{\tilde t}= 2 m_t (A_t + \mu \cot \beta)/(m^2_{\ta} -m^2_{\tb})$  prevents the loop corrections from growing substantially as the soft scalar masses grow in the models we discuss.

\section{Difference between an Intersection Point and a Focus Point}
The result we uncover is not the focus point solution.
The focus point is actually a sub-case of the more the general situation discussed prior in Ref.~\cite{Nath:97} where the soft parameters sit on a hyperbola in the  gaugino-scalar mass plane, i.e. $(M_{1/2},M_0)$  allowing $m^2_{H_u}$ to be either positive or tachyonic owing to cancellations in the RG flow~\cite{Arnowitt:97,Nath:97}.

We now explain in more detail how this solution is different from the focus point solution~\cite{Nath:97,Feng:2000}. The focus point solution to the RGE for $m^2_{H_u}$ occurs when the product $f_{A_0} A^2_0$ is tiny compared to $f_{M_0} M^2_0$ and holds only for small trilinear couplings and gaugino masses. In that case it is the coefficient of $f_{M_0}$ which becomes small and suppresses $m^2_{H_u}$   and allows the scalar soft mass$^2$ in the $(few ~\rm TeV)^2$  range to run to essentially a common focal point~\cite{Feng:2000} in the plane spanned by the running scale and  $m^2_{H_u}$, driving it 
negative,  which as re-emphasized here, is not what happens for an intersection point.  At an intersection point, the trilinears are the same size as the very heavy scalars above the 10 TeV range allowing for the cancellation  between both their RG coefficients.

The intersection point we discuss here is a phenomenon that has not been noticed before.  Our analysis suggests that the intersection point does appear to live within the hyperbolic branch, and is not  the focus point solution sub-case, however, the intersection point  was not noticed until our analysis in this work, as  
in both the analyses of Refs.~\cite{Nath:97,Feng:2000} the largest magnitude of the trilinear couplings never exceeded $\sim 4 ~\rm TeV$; in either case,   
the effects we discuss were not demonstrated.  

The analysis presented here, semi-analytical and numerical,  shows that the intersection point has a major impact on the physics. Namely,  supergravity and string motivated models have  a built in mechanism, the intersection point, that can provide a  consistent, rather natural value of $\mu$, sub-TeV to a few TeV  with  radiative breaking of electroweak symmetry while providing a solution to the cosmological moduli and gravitino problems.
\\

{\it Acknowledgements:}
We collectively thank Tim Cohen, Piyush Kumar, Pran Nath, Brent Nelson, Aaron Pierce, Lian-Tao Wang, and James Wells, and Hai-Bo Yu for comments and conversations following the near completion of this paper.  This research is supported by Department of Energy  grant DE-FG02-95ER40899 and by the Michigan Center for Theoretical Physics, and support of EK and RL by the String Vacuum Project Grant funded through NSF grant PHY/0917807.

\end{document}